\documentclass[xypic,amscd,syntonly,amssymb,verbatim,12pt]{amsart}

\usepackage{graphicx}
\usepackage[all]{xy}
\usepackage{amssymb}
\usepackage{amsmath}
\usepackage[mathscr]{euscript}

\usepackage{epsfig}
\theoremstyle{plain}
\newtheorem{Thm}{Theorem}
\newtheorem{Prop}[Thm]{Proposition}
\newtheorem{Cor}[Thm]{Corollary}
\newtheorem{Lem}[Thm]{Lemma}

 \theoremstyle{definition}
\theoremstyle{remark}

\numberwithin{equation}{section}

\begin{document}
  \title{Branes on $G$-manifolds}

 \author{ ANDR\'{E}S   VI\~{N}A}
\address{Departamento de F\'{i}sica. Universidad de Oviedo.   Avda Calvo
 Sotelo.     33007 Oviedo. Spain. }
 \email{vina@uniovi.es}

 \maketitle
\begin{abstract}
 Let $X$ be  Calabi-Yau manifold acted by a group $G$.
  We give a definition of $G$-equivariance for branes  on $X$, 
 and assign to each equivariant brane an element of the equivariant cohomology of $X$ that can be considered as a charge of the brane.
We prove that the spaces of
strings stretching between equivariant branes support
representations of $G$. This fact allows us to give formulas for   the dimension of some of such spaces, when $X$ is a flag manifold of $G$.

 \end{abstract}

\section {Introduction} \label{S:intro}

 Let $X$ be a compact  K\"ahler $n$-manifold analytically   acted by a Lie group $G$.
 Some objects related  with $X$ admit an ``equivariant" version, 
  when they are equipped with   a $G$-action compatible with  its structure; for example,
  the equivariant vector bundles on $X$.

In the same way, it seems    natural to consider ``equivariant" $B$-branes  on $X$; i.e., $B$-branes
endowed with a $G$-action which lifts the $G$-action on $X$. In
this article we will concern with such branes.  We will associate
them  ``equivariant charges", consider its equivariant cohomology, study the spaces of strings stretching between ``equivariant" branes and the correlation functions for vertex operators for these strings, etc. 

 Besides the Introduction, the article consists of three sections: Equivariant branes, Cohomology of equivariant branes and an Appendix. In the Appendix, we present   detailed proofs
 of two propositions which are stated in Section \ref{S:SectEqSh}.
 The following is a brief explanation  of the key points 
considered in  the first two sections.

  
\smallskip

 {\bf Equivariant branes.} A $D$-brane of type $B$ on
  $X$   can be considered as an  object of the derived category of coherent
sheaves on $X$ (see   monograph \cite{Aspin-et}, which includes a large list of specific references).
 Particular $B$-branes are
  the coherent sheaves.
 There exists a definition of $G$-equivariance for sheaves, which generalizes the one for equivariant vector
   bundles and, obviously, applicable to branes which are coherent sheaves. (In Subsection \ref{SubSectEquivSheav}, we recall this concept). 

 To explain  the extension of that definition to a general brane, we introduce some notations. 
In the definition of $B$-branes is involved the category $\mathfrak{Coh}$ of coherent sheaves on $X$, which is a subcategory of  $\mathfrak{Mod}$,   the category of ${\mathcal O}_X$-modules \cite{Kas-Sch}.
  We put $\mathfrak{Mod}^G$  for the subcategory of $\mathfrak{Mod}$ whose objects are $G$-equivariant sheaves
  and we denote by $\mathfrak{Coh}^G$ the full
  subcategory of $\mathfrak{Mod}^G$ whose objects belong to $\mathfrak{Coh}$.
 Hence, we have the following subcategories of $\mathfrak{Mod}$.
 $$
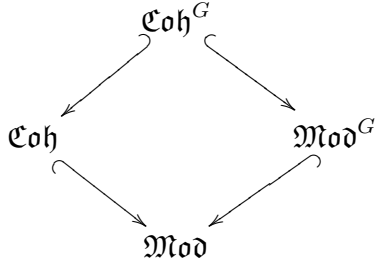
\xymatrix{ & \mathfrak{Coh}^G \ar@{_{(}->}[ld]\ar@{^{(}->}[rd] \\
  \mathfrak{Coh} \ar@{_{(}->}[rd] &{} &  \mathfrak{Mod}^G  \ar@{^{(}->}[ld] \\
 &  \mathfrak{Mod} }
 $$

 As we said, a brane is an object of $D(\mathfrak{Coh})$,
   the bounded derived category of $\mathfrak{Coh}$.  One possible translation   of the concept of equivariance to  more  general  branes,  is to define the $G$-equivariant branes  as the objects of $D(\mathfrak{Coh}^G)$,
   the bounded derived category  of the abelian category $\mathfrak{Coh}^G$. In this article we will adopt this point of view.
	
	In the definition of the spaces of strings between branes are involved the ${\mathit Ext}$ groups. More precisely, let ${\mathcal F}$ and ${\mathcal G}$ be two   general
$B$-branes, then  an open string
between ${\mathcal F}$ and ${\mathcal G}$ with ghost number $k$ is
an element of the Ext group $\mathit{Ext}^k({\mathcal
F},\,{\mathcal G})$
 \cite[Sect. 5.2]{Aspin}. 
	On the other hand,	the space of local operators for
  strings with   ghost number $k$ stretching from ${\mathcal F}$ to ${\mathcal G}$ is 
		(see \cite{Vina})
 \begin{equation}\label{bigoplusk}
 \bigoplus_q H^q\big(X,\,{\mathcal Ext}^k({\mathcal F},\,{\mathcal
 G})\big).
 \end{equation}

    Although $\mathfrak{Coh}^G$ has not enough injectives, the category $\mathfrak{Mod}^G$ is abelian and
  it has  sufficient injectives \cite{Tohoku}.   The existence of ``sufficient injectives" allows us to construct in $\mathfrak{Mod}^G$ the $\mathit{Ext}$ groups,
	and the   ${\mathcal Ext}$ functors.
	That is,
    the space of strings between two objects of $D(\mathfrak{Coh}^G)$
    can be defined by considering them as objects of  $D(\mathfrak{Mod}^G)$, the derived category of $\mathfrak{Mod}^G$. In this way, the  equivariance  of
    branes gives rise to representations of $G$ on the corresponding  spaces of strings   and  on the vertex operators. These results are stated in Propositions \ref{PropArbitraryBranes} and \ref{Repres_Strings}.
		
	When ${\mathcal F}$ and ${\mathcal G}$ are locally free sheaves,   we will prove that the correlation functions of vertex operators for strings between these branes are $G$-invariant (Proposition \ref{Thm:CorrelationInvar}).

The Borel-Weil-Bott theorem permits 	to determine the dimension of the space of strings between branes which are $G$-equivariant locally free sheaves on a flag manifold of $G$; the result is stated in Proposition \ref{P:Bore-Weil-Bott}. When the branes are rank $1$ locally free sheaves, we determine the highest weight of the representation of $G$ supported by the corresponding spaces of vertex operators (Proposition  \ref{Flag}). In this way, the computation of the dimension of these spaces   reduces to the application of the Weyl's dimension formula.



	\smallskip
		\noindent
		{\bf Cohomology of equivariant branes.} By $\bar X$ we denote the homotopy quotient $\bar X:=EG\times_G X$, where $EG$ is the universal bundle of the group $G$. Given an equivariant brane ${\mathcal F}$ on $X$, it admits a lift to an object $\bar{\mathcal F}$ of the derived category of sheaves on $\bar X$. In fact, the pair $(\bar{\mathcal F},\,{\mathcal F})$ determines an object in the equivariant derived category $D_G(X)$, introduced by Bernstein and Lunts in \cite{Be-Lu}. Thus, one can define the equivariant cohomology of the brane ${\mathcal F}$, as the cohomology $H(\bar X,\,\bar{\mathcal F})$.
		
		In the case that the group $G$ is a compact torus $T$, the localization theorems for the equivariant cohomology are applicable to the groups
	 $H^p(\bar X,\,\bar{\mathcal F})$. In this way we give, in Corollary \ref{CorFiniteFixed},  a necessary  condition for two $T$-equivariant branes  of $\mathfrak{Coh}^T$ be equivalent.

     When $X$ is an algebraic variety and under certains hypotheses on the $G$-action, each object of $\mathfrak{Coh}^G$ admits a resolution consisting of $G$-equivariant locally free sheaves \cite{Thomason1}. Thus, it is possible to define $G$-equivariant charges for the branes  of $D(\mathfrak{Coh}^G)$; i.e., elements of the equivariant cohomology $H_G(X)$, which coincide with the usual charges when $G=\{1\}$. 
  For a brane ${\mathcal F}\in D(\mathfrak{Coh}^G)$, we define the equivariant charge $Q^G({\mathcal F})$ as the product of the equivariant Chern character of ${\mathcal F}$ and the equivariant Todd class of $X$.
  
 Some equivariant charges admit interpretations in terms of the index of elliptic operators. 
   For example,  when the equivariant brane is the sheaf ${\mathcal O}(V)$, of sections of the holomorphic vector bundle  $V$, and $G$ acts on $X$ as a group of isometries, then  the Dirac operator $D$ defined on $(\bigwedge T^*X)\otimes V$ is $G$-equivariant, and ${\rm ker}(D)-{\rm coker}(D)$ supports a virtual representation of $G$. The character of this representation is equal to the evaluation of $Q^G(V)$ on $X$ (see Proposition \ref{P:DiracVirtual}).
 
  Let $X$ be a toric manifold  and $T$ be the torus whose action on $X$ defines the toric structure. Given a $T$-equivariant brane on $X$ which is a locally free sheaf ${\mathcal O}(V)$,  applying     the localization formulas of equivariant cohomology, we will evaluate the  equivariant charge $Q^T({\mathcal O}(V))$
   in terms of data associated to the fixed points of the $T$-action on $X$. The result is stated in Proposition \ref{PropFixedPoint}.

  \smallskip

{\it Acknowledgement.} I thank Diego Rodr\'{\i}guez-G\'omez for his useful comments.


\section{Equivariant branes}\label{S:SectEqSh} In this
section, we introduce the category $\mathfrak{Mod}^G$ of
$G$-equivariant ${\mathcal O}$-modules, define a representation on
the cohomology groups of the branes which are objects of
$D(\mathfrak{Coh}^G)$, and characterize the space of strings stretching between some equivariant branes on a flag manifold of $G$.

\subsection{Equivariant sheaves}\label{SubSectEquivSheav}  By ${\mathcal O}_X$, or simply by ${\mathcal O}$,  we denote the
sheaf of regular functions on $X$.
 Let  $\mu:G\times X\to X$  be an {\em analytic} action of a reductive Lie group $G$
on $X$. Essentially,
 a $G$-equivariant structure on the ${\mathcal O}$-module  ${\mathcal H}$ is given by a family
 $\{\lambda_{g,x}\}$ of isomorphisms between the stalks
 \begin{equation}\label{asterisco1}
 \lambda_{g,x}:{\mathcal H}_x  \to{\mathcal H}_{\mu(g,x)},\;\;\;\hbox{for all}\;\;g\in G,\;x\in X,
  \end{equation}
 compatible with the multiplication in $G$ (i. e., satisfying the cocycle condition).

 To formulate the cocycle condition,
 we introduce the following notations
$$m:  G\times G\to  G,\;\;\;m(g_1,\,g_2)=g_1g_2.\;\;\;\;
 b: G\times X\mapsto x\in X,\;\;\; b(g,x)=x.$$
$$p:G\times G\times X\to G\times X,\;\;\; p(g_1,\,g_2\,,x)=(g_2\,,x).$$
Thus, one has the maps $p,$ $ m\times 1_X$ and $1_G\times \mu$
from $G\times G\times X$ to $G\times X$ and the corresponding
functors
\begin{equation}\label{functors0}
\xymatrix{\mathfrak{Mod}({\mathcal O}_X)\ar@/^/[r]^{b^*}
\ar@/_/[r]_{\mu^*} & \mathfrak{Mod}({\mathcal O}_{G\times X})
\ar[r]^{p^*} \ar@/_1pc/[r]_{(1_G\times \mu)^*}
\ar@/^2pc/[r]^{(m\times 1_X)^*} & \mathfrak{Mod}({\mathcal
O}_{G\times G\times X})\,, }
\end{equation}
where an asterisk as superscript is used for denoting  the inverse
image functor between the corresponding  categories, and
$\mathfrak{Mod}({\mathcal O}_Z)$ stands for the category of
${\mathcal O}_Z$-modules. The equalities
$$b\circ(m\times 1_X)=b\circ p,\,\; b\circ(1_G\times\mu)=\mu\circ p,\,\; \mu\circ(1_G\times\mu)=\mu\circ(m\times 1_X)$$
give rise to equalities between the respective compositions of the
functors in (\ref{functors0}).

Let ${\mathcal H}$ be an ${\mathcal O}$-module and $\lambda$ an
isomorphism $\lambda: b^{*}{\mathcal H}\to\mu^{*}{\mathcal H}$,
which satisfies the cocycle condition
 \begin{equation}\label{cocycle2}
(m\times 1_X)^*(\lambda)=(1_G\times\mu)^*(\lambda)\circ
p^*(\lambda).
\end{equation}
 We say that  the pair $({\mathcal H},\,\lambda)$
    is a $G$-equivariant sheaf of ${\mathcal O}$-modules, or simply a $G$-equivariant ${\mathcal O}$-module.

One  defines the category $\mathfrak{Mod}^G$, whose objects
    are the $G$-equivariant  ${\mathcal O}$-modules.  If $({\mathcal H}',\,\lambda')$
    and $({\mathcal H},\,\lambda)$ are objects in this category, a morphism in $\mathfrak{Mod}^G$
    from $({\mathcal H}',\,\lambda')$ to $({\mathcal H},\,\lambda)$ is a morphism  of ${\mathcal O}$-modules $f:{\mathcal H}'\to{\mathcal H}$,
   such that $\lambda b^*(f)=\mu^*(f)\lambda'$.

Given $g\in G$, we  put $L_g$ for the map defined by
  $x\in X\mapsto \mu(g,x)=gx\in X$.  Since the $G$-action on $X$ is analytic,  the composition with $\mu$ defines a morphism of sheaves
 of rings ${\mathcal O}_X\to{\mathcal O}_{G\times X}$. In particular,  given an open subset $U\subset X$ and $g\in G$,
  the map
  $$h\in{\mathcal O}(U)\mapsto h\circ L_{g^{-1}}\in {\mathcal O}(gU)$$
   determines a ring isomorphism ${\mathcal O}(U)\to {\mathcal O}(gU).$
In other terms, we have an isomorphism of sheaves of rings
 \begin{equation}\label{O-LgO}
 {\mathcal O}\to (L_{g^{-1}})_*{\mathcal O}.
 \end{equation}



We have the following proposition.

\begin{Prop}\label{P:Propiso}
 Let $({\mathcal H},\,\lambda)$ be a $G$-equivariant ${\mathcal O}$-module and $g$ an element of $G$,
 then $\lambda$ determines  an isomorphism  of ${\mathcal O}$-modules
 $$\lambda_g:{\mathcal H}\to (L_{g^{-1}})_*{\mathcal H}, $$
 where the ${\mathcal O}$-structure of $(L_{g^{-1}})_*{\mathcal H}$ is defined through the isomorphism (\ref{O-LgO}).
  \end{Prop}

The image of
$\sigma_U\in{\mathcal H}(U)$ by the isomorphism
${\mathcal H}(U) \stackrel{\sim}{\longrightarrow} {\mathcal
 H}(gU)$
 will be denoted
$g\cdot\sigma_U$.

A consequence of the cocycle condition is  the following proposition. 
\begin{Prop}\label{enumerate}
For each $g\in G$,
\begin{enumerate}
\item $\lambda_h\circ\lambda_g=\lambda_{hg}.$ \item The map
$g\mapsto \Hat{\lambda}_g:=\lambda_g(X)$ is a group homomorphism
from $G$ to the group of automorphisms of the complex vector space
${\mathcal H}(X)$.
\end{enumerate}
\end{Prop}

Although the results stated in Propositions \ref{P:Propiso} and  \ref{enumerate} are easy to understand, we give detailed proofs of these propositions   in Appendix.



In summary,  the cocycle condition for $({\mathcal H},\,\lambda)$
gives rise to a representation of $G$ in the space
$H^0(X,\,{\mathcal H})$.


The following step will be to define a representation in
the cohomologies $H^i(X,\,{\mathcal G})$, when ${\mathcal G}$ is an object of   $D(\mathfrak{Coh}^G)$.



\begin{Prop}\label{Proprephcoho}
 If ${\mathcal G}$ is a brane of the category $D(\mathfrak{Coh}^G)$,
  then for
 each $i$ the cohomology group $H^i(X,\,{\mathcal G})$ supports a
 representation of $G$ induced by the $G$-structure of ${\mathcal G}$. When ${\mathcal G}$ is an object of $\mathfrak{Coh}^G$  the
 representation on $H^0(X,\,{\mathcal G})$ is the one of
 Proposition
 \ref{enumerate} (2).
  \end{Prop}
{\it Proof.} As the category $\mathfrak{Mod}^G$ has enough injectives \cite{Tohoku},   following the well-known procedure, it is possible to construct an
equivariant Cartan-Eilenberg resolution ${\mathcal
J}^{\bullet\bullet}$ of ${\mathcal G}^{\bullet}$ in
$\mathfrak{Mod}^G$ \cite[Thm. 10.45]{R}.  Then
${\mathcal G}^{\bullet}$ is quasi-isomorphic to the total complex
${\mathcal I}^{\bullet}=({\rm Tot}({\mathcal J}^{\bullet}),\,\partial^{\bullet})$, a complex of in $\mathfrak{Mod}^G$ consisting of injective objects.

 By the second item in Proposition  \ref{enumerate}, the space the
${\mathcal I}^{i}(X)$ carries the representation $\rho^i$ of $G$.
 Since the diagrams
\begin{equation}\label{DiagExact2}
    \xymatrix{  {\mathcal I^i}(X) \ar[d]_{\rho^i_g}\ar[r]^{{\partial}^i(X)} & {\mathcal I}^{i+1}(X)\ar[d]^{\rho_g^{i+1}} \\
   {\mathcal I}^i(X)\ar[r]^{{\partial}^i(X)} & {\mathcal I}^{i+1}(X) }
 \end{equation}
   are commutative, one has a representation of $G$ on each cohomology group $h^i({\mathcal I}^{\bullet }(X))$ of
 the complex ${\mathcal I}^{\bullet}(X)$. That is, a representation on the cohomology $H^i(X,\,{\mathcal G})$. Thus, we have the proposition.
  \qed

\smallskip

If in the statement of Proposition \ref{Proprephcoho}  ${\mathcal G}$ is a locally free ${\mathcal O}$-module, the representation on $H^i(X,\,{\mathcal G})$ can be constructed by means
of the Dolbeault resolution.
Let ${\mathcal G}={\mathcal O}(V)$ be the sheaf   of germs of sections of the holomorphic vector bundle $V$. We put ${\mathcal A}^{0,q}$ for the sheaf of germs of holomorphic differential forms on $X$ of type $(0,q)$. Since ${\mathcal O}(V)$ is a flat ${\mathcal O}$-module, the tensor product of this module by the Dolbeault resolution of ${\mathcal O}$ gives
 rise to the following fine resolution of ${\mathcal O}(V)$
  \begin{equation}\label{resolutionOmega(V)}
0\to{\mathcal O} (V){\longrightarrow}{\mathcal
A}^{0,0}(V)\overset{1\otimes\bar\partial}
{\longrightarrow}{\mathcal
A}^{0,1}(V)\overset{1\otimes\bar\partial}
{\longrightarrow}{\mathcal A}^{0,2}(V){\longrightarrow}\dots
\end{equation}
 where ${\mathcal A}^{0,q}(V):={\mathcal O}(V)\otimes_{\mathcal O}{\mathcal A}^{0,q}.$
   Thus,
\begin{equation}\label{Hq(X}
 H^q(X,\,{\mathcal O}(V))=h^q(A^{0,\bullet}(V)),
\end{equation}
 where $A^{0,q}=\Gamma(X,\,{\mathcal A}^{0,q}(V))$.

Now, we use the fact that $V$ is a $G$-equivariant vector bundle. If  $\sigma$ a holomorphic section of   $V$ and $\omega$ a $(0,q)$-form on $X$, we put
\begin{equation}\label{gcdot(}
g\cdot (\sigma\otimes \omega):=(g\cdot\sigma)\otimes L_{g^{-1}}^*\omega,
 \end{equation}
  $L_{g^{-1}}^*\omega$ being the pullback of $\omega$ by the diffeomorphism
  $L_{g^{-1}}$ and $g\cdot\sigma$ the section defined as just after
  Proposition \ref{P:Propiso}.

  If $f$ is a holomorphic function on $X$, then
 $$g\cdot(f\sigma\otimes\omega)=g\cdot(\sigma\otimes f\omega).$$
  On the other hand, since the $G$ acts analytically on $X$,
  $1\otimes\bar\partial$  commutes with the $G$-action (\ref{gcdot(}). So, we have  a representation of $G$
  on $H^q(X,\,{\mathcal O}(V))$, which is equivalent to the one of Proposition \ref{Proprephcoho}.

\smallskip


\subsection{Vertex operators.} In this subsection, we consider the spaces of vertex operators  for
 strings stretching between equivariant branes.

We put ${\mathcal Hom}(\,.\,,\,.\,)$ for the sheaf functor Hom of the category $\mathfrak{Mod}$
(see \cite[page 87]{Kas-Sch})
$${\mathcal Hom}(\,.\,,\,.\,): \mathfrak{Mod}^{\rm op}\times\mathfrak{Mod}\to \mathfrak{Sh},$$ 
where $\mathfrak{Sh}$ is the category of sheaves of abelian groups over $X$.

Let $({\mathcal F},\,\gamma)$, $({\mathcal G},\,\beta)$ be
$G$-equivariant ${\mathcal O}$-modules.
We set ${\mathcal K}:={\mathcal Hom}({\mathcal F},\,{\mathcal G})$ for the sheaf of homomorphisms
from ${\mathcal F}$ to ${\mathcal G}$.
 Given an open  subset $U\subset X$, a section $\Phi\in {\mathcal K}(U)={\rm
Hom}_{{\mathcal O}(U)}({\mathcal F}(U),\,{\mathcal G}(U))$ and
$g\in G$, we define $g\cdot \Phi\in {\mathcal K}(gU)$ as follows:

Let $\tau\in {\mathcal F}(gU)$ be a section of ${\mathcal F}$ on
$gU$, then $g^{-1}\cdot\tau\in {\mathcal F}(U)$,  with the
notation introduced below   Proposition \ref{P:Propiso}. Thus,
$\Phi(g^{-1}\cdot\tau)\in {\mathcal G}(U)$ and
$g\cdot\Phi(g^{-1}\cdot\tau)\in {\mathcal G}(gU)$.
 We put
 \begin{equation}\label{gcdotP}
 (g\cdot\Phi)(\tau):=g\cdot(\Phi(g^{-1}\cdot\tau)).
 \end{equation}

 So, we have constructed an isomorphism
 \begin{equation}\label{Etau}
   \eta_g(U)
   :{\mathcal K}(U)\longrightarrow
{\mathcal K}(gU),\;\; \;\;  \Phi\mapsto g\cdot \Phi.
\end{equation}
Moreover,
\begin{equation}\label{EquatioEta}
 \eta_h(gU)\circ \eta_g(U)=\eta_{hg}(U).
 \end{equation}
 Hence, the isomorphisms
 $\{\eta_g(X)\}_g$
 define a representation of $G$ on the space ${\mathcal K}(X)$. Thus, one has the following proposition.

\begin{Prop}\label{PropEta}
Let $({\mathcal F},\,\gamma)$, $({\mathcal G},\,\beta)$ be
$G$-equivariant ${\mathcal O}$-modules,  then
\begin{enumerate}
\item The isomorphisms (\ref{Etau}) convert the ${\mathcal O}$-module
$${\mathcal K}={\mathcal Hom}({\mathcal F},\,{\mathcal G})$$
in a $G$-equivariant ${\mathcal O}$-module.
\item For $g\in G$ and $\Psi\in{\mathcal K}(X)={\rm Hom}_{\mathfrak{Mod}}({\mathcal
F},\,{\mathcal G})$, we put
$g\cdot\Psi$ for the element of ${\mathcal K}(X)$ defined by
$$(g\cdot\Psi)_U(\sigma):= g\cdot(\Psi_{g^{-1}U}(g^{-1}\cdot\sigma)),$$
where $U$ is an open subset of $X$ and $\sigma\in{\mathcal F}(U)$. The correspondence $\Psi\to g\cdot\Psi$
  defines a representation of $G$ in vector space ${\mathcal K}(X)$.
\end{enumerate}
\end{Prop}

Given ${\mathcal G}^{\bullet}$ a complex in the category $\mathfrak{Coh}^G$, we consider ${\mathcal I}^{\bullet}$ the complex consting of injective objects in $\mathfrak{Mod}^G$ described  in the proof of Proposition \ref{Proprephcoho}. 
If ${\mathcal F}^{\bullet}$ is another complex in
$\mathfrak{Mod}^G$, we put
$${\mathcal C}^n:=\prod_a{\mathcal Hom}({\mathcal F^a,\,{\mathcal
I}^{a+n}}).$$ By Proposition \ref{PropEta}, each ${\mathcal
Hom}({\mathcal F^a,\,{\mathcal I}^{a+n}})$ is an object of
$\mathfrak{Mod}^G$.


On the other hand,
 the coboundary operator $\delta^n:{\mathcal C}^n\to{\mathcal C}^{n+1}$ is defined  by (\cite[page 17]{Iversen})
$$\delta^n(\Psi_a)=\big(\partial^{a+n}\circ\Psi_a+(-1)^{n+1}\Psi_{n+1}\circ\partial ^a  \big),$$
where
$$\Psi_a\in {\mathcal Hom}({\mathcal F^a,\,{\mathcal
I}^{a+n}}).$$
Since the   $\partial$'s are $G$-operators, so is
 $\delta^n$. Thus, $({\mathcal
 C}^{\bullet},\,\delta^{\bullet})$ is a complex in the category
 $\mathfrak{Mod}^G$.  The complex  ${\mathcal
 C}^{\bullet}$ is   usually denoted by ${\mathcal
 Hom}^{\bullet}\big({\mathcal F}^{\bullet},\,{\mathcal I}^{\bullet}
 \big).$ By definition, ${\mathcal Ext}^p({\mathcal F}^{\bullet},\,{\mathcal
 G}^{\bullet})$ is the cohomology object
 $h^p({\mathcal
 Hom}^{\bullet}\big({\mathcal F}^{\bullet},\,{\mathcal
 I}^{\bullet})).$

 As ${\mathcal
 Hom}^{\bullet}\big({\mathcal F}^{\bullet},\,{\mathcal I}^{\bullet}
 \big)$ is a complex in the abelian category $\mathfrak{Mod}^G$,
 its cohomologies are also in $\mathfrak{Mod}^G$; i. e. are
 $G$-equivariant sheaves. Hence, from Proposition
 \ref{Proprephcoho}, it follows the following proposition.
 \begin{Prop}\label{PropArbitraryBranes}  Let ${\mathcal F}$ and ${\mathcal G}$ be branes in the derived category
 $D(\mathfrak{Coh}^G)$. 
Then the $G$-structures of ${\mathcal F}$ and ${\mathcal G}$ induce on the space of vertex operators
  $$H^q(X,\,{\mathcal Ext}^p({\mathcal F},\,{\mathcal G}))$$
a representation of $G$.
 \end{Prop}

The spectral sequence of the double complex,
  $$E^{p,q}=H^p(X,\,{\mathcal Ext}^q({\mathcal F},\,{\mathcal G})),$$
  converges to the space of strings ${\mathit Ext}^{p+q}({\mathcal F},\,{\mathcal G})$. Then from Proposition \ref{PropArbitraryBranes}, we conclude:

  \begin{Prop}\label{Repres_Strings}
   If ${\mathcal F}$ and ${\mathcal G}$ are branes of the derived category
  $D(\mathfrak{Coh}^G)$,
   then the space ${\mathit Ext}^k ({\mathcal F},\,{\mathcal G})$ of strings between ${\mathcal F}$ and ${\mathcal G}$ with ghost number $k$
   supports a representation of $G$ induced by the $G$-structures of ${\mathcal F}$ and ${\mathcal G}$.
  \end{Prop}


When ${\mathcal F}$ and ${\mathcal G}$ are locally free ${\mathcal
O}$-modules the representation stated in the previous proposition can
be formulated in terms of differential forms.

First of all, one has the following resolution for ${\mathcal F}$
consisting of locally free ${\mathcal O}$-modules
 $$\dots\to 0\to 0\to 0\to{\mathcal F}\overset{1}{\to}{\mathcal F}\to
 0.$$
 By \cite[Proposition 6.5, page 234]{Hart}, ${\mathcal Ext}^k({\mathcal
 F},\,{\mathcal G})=0$, for $k\ne 0$. Hence, when ${\mathcal F}$ is a locally free module we consider only the
 spaces of vertex operators $H^q(X,\,{\mathcal Hom}({\mathcal
 F},\,{\mathcal G}))$. Furthermore, in this case,
\begin{equation}\label{cuerdasVectorbundles}
{\mathit Ext}^k({\mathcal F},\,{\mathcal G})=H^k(X,\,{\mathcal Hom}({\mathcal F},\,{\mathcal G})).
  \end{equation}

Let  $V_1$ and $V_2$ be two   holomorphic $G$-equivariant vector
bundles on $X$. We denote by $V$
  the holomorphic  vector bundle   ${\rm Hom} (V_1,\,V_2)$. Given $\psi$ is an element
  of $\Gamma(X,\,{\mathcal O}(V))$, putting
\begin{equation}\label{gcdotAux}
(g\cdot\psi)(-)=g\cdot(\psi(g^{-1}\cdot (-)))
 \end{equation}
 we
define a representation of $G$ on $\Gamma(X,\,{\mathcal O}(V))$; it
is the one stated in item 2 of Proposition \ref{PropEta}. More
explicit, given $x\in X$, the homomorphism $(g\cdot
\psi)_x:V_{1x}\to V_{2x}$ is the composition
\begin{equation}\label{composition}
\lambda^2_{g,g^{-1}x}\circ\psi_{g^{-1}x}\circ\lambda^1_{g^{-1},x},
 \end{equation}
where the $\lambda $'s are the isomorphisms (\ref{asterisco1}); that is, $\lambda^i_{g,y}:(V_i)_y\to (V_i)_{gy}$.

As   ${\mathcal Hom}({\mathcal O}(V_1),\,{\mathcal
O}(V_2))={\mathcal O}({\rm Hom}(V_1,\,V_2))$,  then (\ref{Hq(X})
applied to $V={\rm Hom}(V_1,\,V_2)$ gives
 \begin{equation}\label{Hq(X1}
H^q(X,\,{\mathcal Hom}({\mathcal O}(V_1),\,{\mathcal O}(V_2)))
= h^{q}(A^{0,\bullet}({\rm Hom}(V_1,\,V_2))).
  \end{equation}
 As in (\ref{gcdot(}), from the $G$-action
 (\ref{gcdotAux}),  we can construct the corresponding representation
 on the space (\ref{Hq(X1}). Thus, given
 $$\psi\in\Gamma(X,\,{\mathcal Hom}({\mathcal
 O}(V_1),\,{\mathcal O}(V_2)))$$
  and the differential form $\omega$ of type
 $(0,q)$, we have
 \begin{equation}\label{gcdotLOCALLY}
 g\cdot(\psi\otimes\omega)=g\cdot
 \psi\otimes L^*_{g^{-1}}\omega.
  \end{equation}
Thus, Proposition \ref{PropArbitraryBranes} adopts the following
form when the branes are locally free sheaves.
\begin{Prop}\label{P:equivariant-Llocally} Let $V_1$, $V_2$ be $G$-equivariant holomorphic vector
bundles on $X$. Then the $G$-action (\ref{gcdotLOCALLY}) induces a
representation of $G$ on the space of vertex operators
$$H^q(X,\,{\mathcal Hom}({\mathcal O}(V_1),\,{\mathcal
O}(V_2))).$$
\end{Prop}  

\smallskip

\noindent
{\bf Vertex operators on flag varieties.} When $X$ is a flag manifold, the dimension of the spaces of vertex operators mentioned in
 Proposition \ref{P:equivariant-Llocally} can be determined, in some particular cases, by means of the Borel-Weil-Bott theorem and the Weyl's dimension formula.
 
Let $G_{\mathbb C}$   be a connected complex semi-simple Lie group and $Q\subset G_{\mathbb C}$  a parabolic subgroup of $G_{\mathbb C}$. Then the flag manifold  $X=G_{\mathbb C}/Q$ is a compact homogeneous simply connected K\"ahler
  algebraic variety (see for example \cite{Borel, W1}).  Moreover, every  compact homogeneous simply connected K\"ahler manifold is isomorphic to such a quotient.  In \cite{Grant}, Grantcharov showed several examples
  flag manifolds which are also Calabi-Yau.

The holomorphic equivariant vector bundles over $X$ are related with the holomorphic representations of $G_{\mathbb C}$. 
 Let $V\to X$ be a holomorphic vector bundle endowed with a holomorphic $G_{\mathbb C}$-action by bundle maps, that lies over the action on $X$. Denoting with $x_0\in X$ the class $eQ$,  then  stabilizer of $x_0$ acts on the the fibre $V_{x_0}$. Hence, there is a holomorphic representation 
$\xi$ 
of $Q$ on ${\mathsf V}$, the standard fibre of $V$.
   Moreover, one has an isomorphism of holomorphic bundles $V\simeq  G_{\mathbb C}\times_{\xi} {\mathsf V}$. 
Conversely, equivariant holomorphic bundles on $X$ can be constructed from representations.

 We denote by $G$ a compact real form of $G_{\mathbb C}$. 
 As it is  known, the finite dimensional representations of $G$ are in bijective correspondence with the holomorphic representations of finite dimension of $G_{\mathbb C}$.
 We put $L:=G\cap Q$,   then one can identify $G/L$ with  $X$. Moreover, each irreducible representation of  $L$ on a complex space induces a unique extension to one holomorphic representation of $Q$. So, an irreducible representation $\alpha$ of $L$ on the complex vector space ${\mathsf V}$ gives rise to one homogeneous holomorphic vector bundle 
$$V(\alpha):=G_{\mathbb C}\times_Q{\mathsf V},$$
on $X$. 
It is very easy to check that $V(\alpha\oplus\beta)=V(\alpha)\oplus V(\beta),$ and $V(\alpha^*)=(V(\alpha))^*.$

Given $\alpha$, $\beta$ irreducible representations of $L$ the representation tensor product $\alpha^*\otimes\beta$ can be written as direct sum of irreducible representations
\begin{equation}\label{Littlewood-Richard}
\alpha\otimes\beta=\bigoplus_{\nu} m^{\nu}\nu,
 \end{equation}
where $\nu$ is an irreducible representation of $L$ and the $m^{\nu}\in{\mathbb Z}_{\geq 0}$ are the corresponding Littlewood-Richardson coefficients.  

Hence, the bundle of homomorphisms (over the identity) from $V(\alpha)$ to $V(\beta)$ can be written as
\begin{equation}\label{HomValpha}
 \mathit{Hom}(V(\alpha),\,V(\beta))= V(\alpha)^*\otimes V(\beta)= V(\alpha^*\otimes \beta)=\oplus_{\nu}m^{\nu}V(\nu).
 \end{equation}
 Since ${\mathcal O}(V(\alpha))$ is a locally free ${\mathcal O}$-module,   it follows from (\ref{HomValpha}) together with (\ref{cuerdasVectorbundles}) that  the space of string with ghost number $k$ stretching between ${\mathcal O}(V(\alpha))$ and ${\mathcal O}(V(\beta))$
$$\mathit{Ext}^k\big({\mathcal O}(V(\alpha)),\,{\mathcal O}(V(\beta))\big)=\bigoplus_{\nu}m^{\nu}H^k(X,\,{\mathcal O}(V(\nu))).$$
 
 \smallskip
 {\it Notations.}                                                                                
We put
${\mathfrak g}:={\rm Lie}(G_{\mathbb C})$ and ${\mathfrak g}_0$ for ${\rm Lie}(G)$ for the Lie algebras of the corresponding groups. Let $c$ be the complex conjugation in ${ \mathfrak g }$ with respect to  ${ \mathfrak g }_0.$
 We fix a maximal abelian subalgebra ${\mathfrak h}_0$ of ${\mathfrak g}_0$ and   denote by ${\mathfrak h}$ its complexification with respect to $c$. Let $\Delta$ be the set of roots of the Cartan subalgebra ${\mathfrak h}$ in ${\mathfrak g}$. 
 
  The system  of positive roots $\Delta^{+}$ for the pair $({\mathfrak g},\,{\mathfrak h})$ is chosen so that
   $${\rm Lie}(Q)={\mathfrak q}={\mathfrak h}\oplus\bigoplus_{\gamma\in\Phi}{\mathfrak g}^{-\gamma},$$
   where $\Phi$  is a subset of $\Delta$ containing $\Delta^+$ and   ${\mathfrak g}^{-\gamma}$ being the root space associated to $\gamma$.
   We denote by  $\rho$   the half sum of positive roots,
  write $\Lambda$ for the weight lattice and put $\Lambda^+$ for the set of dominant weights.

\smallskip

$\zeta(\alpha)$ denotes the highest weight of the irreducible representation $\alpha$ of $L$.
  By the Borel-Weil-Bott theorem  \cite{Gr-Sch},   $H^k(X,\,{\mathcal O}(V(\nu))=0,$ if $\zeta(\nu)+\rho$ is singular. When $\zeta(\nu)+\rho$ is regular, let $w_{\nu}$ the unique element of the Weyl group    ${\mathbf W}$ such that  $w_{\nu}(\zeta(\nu)+\rho)$ is dominant and define
 $$l({\nu}):=\#\{\alpha\in\Delta^+\,|\, (\zeta(\nu)+\rho,\,\alpha)<0  \},$$
where the bilinear product $(\,.\,,\,.\,)$ is defined by the Killing form.
 The following proposition is consequence of Borel-Weil-Bott theorem.

 \begin{Prop}\label{P:Bore-Weil-Bott}
If $\alpha$ and $\beta$ are irreducible representations of $L$, then the dimension of the space of strings $\mathit{Ext}^k\big({\mathcal O}(V(\alpha)),\,{\mathcal O}(V(\beta))\big)$ is
$$\sum_{\tau}m^{\tau}\delta_{k,l(\tau)}\,{\rm dim}\,\xi_{\tau},$$
where $\tau$ runs over the irreducible representations of $L$, such that $\zeta(\tau)+\rho$ is a regular weight,  $\xi_{\tau}$ being the irreducible representation of $L$ with highest weight $w_{\tau}(\zeta(\tau)+\rho)-\rho$ and the $m^{\tau}$'s the Littlewood-Richardson coefficients defined in (\ref{Littlewood-Richard}).
 \end{Prop}

When the representations $\alpha$ and $\beta$ are unidimensional the statement of Proposition \ref{P:Bore-Weil-Bott} admits a simpler formulation.
 
Given $\upsilon\in \Lambda$, it determines a holomorphic character ${\rm e}^{\upsilon}:Q\to {\mathbb C}^{\times}$. So, one can construct the fiber product $L_{\upsilon}=G_{\mathbb C}\times_{Q}{\mathbb C}$, which is a holomorphic $G_{\mathbb C}$-equivariant line bundle over $X$.
If $\lambda,\mu\in \Lambda$, then
$$L_{-\mu+\lambda}=L_{\mu}^*\otimes L_{\lambda}= \mathit{Hom}(L_{\mu},\,L_{\lambda}).$$
Hence, we have isomorphisms  among the following sheaves of holomorphic sections
$${\mathcal Hom}({\mathcal O}(L_{\mu}),\, {\mathcal O}(L_{\lambda}))= {\mathcal O}\big(\mathit{Hom}(L_{\mu},\,L_{\lambda}) \big)
= {\mathcal O}(L_{-\mu+\lambda}).$$

 We put   $\xi:=-\mu +\lambda$ for the weight that determines the line bundle $L_{\mu}^*\otimes L_{\lambda}$, and 
 $$p(\xi):=\#\{\alpha\in\Delta^+\,|\, (\xi+\rho,\,\alpha)<0  \}.$$
 A direct application of  the Borel-Weil-Bott theorem  to the weight  $\xi$  proves the following proposition, which characterizes some spaces of vertex operators relative to the flag manifold $X=G_{\mathbb C}/Q$.

\begin{Prop}\label{Flag} 
If $\xi+\rho$ is either singular or $p\ne p(\xi)$, then the space of vertex operators 
$H^p(X,\, {\mathcal Hom}({\mathcal O}(L_{\mu}),\, {\mathcal O}(L_{\lambda})) )$
is zero. Otherwise, the   representation of $G$ in that space
 is the irreducible representation with highest weight is  $w(\xi+\rho)-\rho$, where $w$ is the element of the Weyl group such
 that $w(\xi+\rho)$ is dominant.
\end{Prop}

 As ${\mathcal O}(L_{\mu})$ is a locally free ${\mathcal O}$-module, the space of strings from the brane ${\mathcal O}(L_{\mu})$ to
${\mathcal O}(L_{\lambda})$ with ghost number $k$ is
 $$ {\mathit Ext}^k({\mathcal O}(L_{\mu}),\, {\mathcal O}(L_{\lambda}))=
 H^k(X,\, {\mathcal Hom}({\mathcal O}(L_{\mu}),\, {\mathcal O}(L_{\lambda}))).$$
 Thus, taking into account the proposition, 
the Weyl dimension formula permits us to determine the dimension of these spaces.


\smallskip

\subsection{Correlation functions.} Let us assume that $X$ is a
projective Calabi-Yau variety of dimension $n$. For $j=0,\dots,
k$, let ${\mathcal F}_j$ be 
a general brane. Given  $a_j\in
H^{q_j}(X,\,{\mathcal Ext}^{p_j}({\mathcal F}_{j-1},\,{\mathcal
F}_j))$,
 when ${\mathcal F}_k={\mathcal F}_0$ and $\sum(q_j+p_j)=n$, one
 can define the correlation function $\langle a_1\dots a_k\rangle$
 of the vertex operators $a_j$'s (see \cite{Vina}).

 If the ${\mathcal F}_j$ are locally free sheaves, i.e. ${\mathcal F}_j={\mathcal
 O}(V_j)$, with $V_j$ holomorphic vector bundle, the correlation
 function  can be calculated as follows.
 Given
 $$a_j\in H^{q_j}(X,\,{\mathcal Hom}({\mathcal
 O}(V_{j-1}),\,{\mathcal O}(V_j)))=H^{0.q_j}_{\bar\partial}(X, {\rm Hom}(V_{j-1},\,V_j)),$$
 $a_j$ will be the class of a form $\psi^j\otimes\omega^j$, with $\psi^j$
 a holomorphic section of ${\rm Hom}(V_{j-1},\,V_j)$ and
 $\omega^j$ a differential  form on $X$ of type $(0,\,q_j)$. Then
 \cite{Aspin, Vina}
\begin{equation}\label{CorrelationLocally}
\langle a_1\dots a_k\rangle=\int_X{\rm
tr}(\psi^k\circ\dots\circ\psi^1)\,\omega^k\wedge\dots\wedge\omega^1\wedge\Omega,
 \end{equation}
where $\Omega$ is  an fixed $n$-holomorphic form on $X$ which vanishes
nowhere.

Next, let us assume that the ${\mathcal O}$-modules ${\mathcal
F}_j$ are $G$-equivariant, and we denote  by $\lambda^j$ the isomorphism that defines the $G$-structure on ${\mathcal F}_j$. Given $g\in G$, by (\ref{gcdotLOCALLY})
\begin{equation}\label{Correlationg}
\langle g\cdot a_1\dots g\cdot a_k\rangle=\int_X{\rm
tr}(g\cdot\psi^k\circ\dots\circ
g\cdot\psi^1)\,L^*_{g^{-1}}\omega^k\wedge\dots\wedge
L^*_{g^{-1}}\omega^1\wedge\Omega.
 \end{equation}
 From (\ref{composition}) together with the cocycle condition, it follows
 $$(g\cdot\psi^j)_x\circ(g\cdot\psi^{j-1})_x=\lambda^j_{g,g^{-1}x}\circ\psi^j_{g^{-1}x}\circ \psi^{j-1}_{g^{-1}x}\circ\lambda^{j-2}_{g^{-1},x}.$$
 As $\lambda^0=\lambda^k$
\begin{align}
{\rm tr}(g\cdot\psi^k\circ\dots\circ g\cdot\psi^1)(x)&={\rm
tr}(\lambda^0_{g,g^{-1}x}\circ\psi^k_{g^{-1}x}\circ\dots\circ\psi^1_{g^{-1}x}\circ\lambda^0_{g^{-1},x})
 \notag \\
&={\rm tr}(\psi^k\circ\dots\circ \psi^1)(g^{-1}\cdot x). \notag
\end{align}

Since the $G$ action on $X$ is analytic,
$L^*_{g^{-1}}\Omega=\Omega$; so, the integrals
(\ref{CorrelationLocally}) and  (\ref{Correlationg}) are equal;
that is,
\begin{equation}\label{langecdota}
\langle g\cdot a_1\dots g\cdot a_k\rangle=\langle a_1\dots
a_k\rangle.
\end{equation}
Thus, we have the following proposition.

\begin{Prop}\label{Thm:CorrelationInvar}
Let ${\mathcal F}$ and ${\mathcal G}$ be   $G$-equivariant locally free ${\mathcal O}$-modules, then the correlation functions associated to vertex operators for strings between ${\mathcal F}$ to ${\mathcal G} $ are $G$-invariant.
\end{Prop}


\section{Cohomology of equivariant branes}\label{S:equivariant}

Because of the equivariance, the branes in  $D(\mathfrak{Coh}^G)$ admit a lift to  branes on the homotopy quotient of $X$, and so one can   define the corresponding equivariant cohomology for these objects. In this section, we will consider that equivariant cohomology.


Let $G$ be a compact Lie group. By the Peter-Weyl theorem $G$ is a
closed subgroup of $GL(N,\,{\mathbb R})$ for some $N$. The limit
as $n\to\infty$ of  Stiefel manifold of $N$-frames in ${\mathbb
R}^{N+n}$ is a smooth model for the universal $G$-bundle.
 We will adopt this model and denote by  
  $EG\overset{q}{\to} BG:=EG/G$ the corresponding $G$-fibration.

 We put $\bar X:=(EG\times X)/G$ the homotopy quotient
of $X$ by $G$ and set $\tau$ and $\pi$ for the projections
 $X\overset{\tau}{\longleftarrow} EG\times X\overset{\pi}{\longrightarrow} \bar X.$
 Let $X\to\bar X=EG\times_G X\overset{\nu}{\to}BG$ be the fibration
constructed from the action of $G$ on $X$, and let $p$ denote the composition
$EG\times X \overset{proj.} {\longrightarrow} EG\overset{q}{\to} BG$. One has the following commutative diagram 
$$ \xymatrix{ \bar X \ar[rd]_{\nu} & EG\times X \ar[l]_{\pi} \ar[d]^{p} \ar[r]^{\tau} & X\\
 {}&  BG & {}
 }$$
 
Let ${\mathcal H}$ be a $G$-equivariant coherent sheaf  on $X$, then the inverse image $\tilde{\mathcal H}:=\tau^*{\mathcal H}$ is $G$-equivariant  coherent sheaf on $EG\times X$. Denoting  by $\tilde L_g$ the obvious multiplication by $g\in G$ in $EG\times X$, we put  $\tilde\lambda_g$ by the isomorphism $\tilde\lambda_g:\tilde{\mathcal H}\to(\tilde L_{g^{-1}})_*\tilde{\mathcal H}$ determined by the equivariance of $\tilde{\mathcal H}$ (see Proposition \ref{P:Propiso}).

 On the other hand, if $V$ is an open set of $\bar X$, $\pi_*\tilde{\mathcal H}(V)=\tilde{\mathcal H}(W)$, where $W$ is the $G$-invariant subset $\pi^{-1}(V)$.  Thus, there is a $G$-action on the sheaf $\pi_*\tilde{\mathcal H}$ defined on 
$\pi_*\tilde{\mathcal H}(V)$ by the isomorphism $\tilde\lambda_{g,W}$.
 We put $\bar{\mathcal H}$ for denoting the subsheaf of $\pi_*\tilde{\mathcal H}$  defined by $G$-invariant sections of $\pi_*\tilde{\mathcal H}$.
Then the sheaves $\tau^*{\mathcal H}$ and $\pi^*\bar{\mathcal H}$ are isomorphic (see \cite[page 3]{Be-Lu}) and denote by $h$ the isomorphism
 $h:\tau^*{\mathcal H}\to\pi^*\bar{\mathcal H}$.

For any open not empty subset $ U\subset BG$, one has $\tau p^{-1}(U)=X$  and $\nu^{-1}(U)=\pi p^{-1}(U)$.   
 \begin{align} \label{H0nu}
\Gamma(\nu^{-1}(U),\,\bar{\mathcal H}) &=
\bar{\mathcal H}(\pi p^{-1}(U))=\pi^*\bar{\mathcal H}(p^{-1}(U)) \\
&\simeq\tau^*{\mathcal H}(p^{-1}(U))=\Gamma(X,\,{\mathcal H}), \notag
 \end{align}
where the isomorphism is determined by $h$. We have proved the following proposition.
\begin{Prop}\label{P:AUXforH}
Given ${\mathcal H}$   a $G$-equivariant coherent sheaf  on $X$,   it determines a sheaf $\bar{\mathcal H}$ on $\bar X$ and an isomorphism
$\tau^*{\mathcal H}\simeq\pi^*\bar{\mathcal H}.$ Furthermore, $\nu_*\bar{\mathcal H}$ is the constant sheaf $\Gamma(X,\,{\mathcal H})$. 
\end{Prop}

Given   ${\mathcal F}$ an object of $D(\mathfrak{Coh}^G)$, i.e., a $G$-equivariant brane, according to the preceding remarks, it defines a object
$\bar{\mathcal F}$ of the derived category $D(\mathfrak{Sh}(\bar X))$, of abelian sheaves on $\bar X$, and an isomorphism $f$ of   the category of $D(\mathfrak{Sh}  
  ({EG\times X}))$   from $\tau^*({\mathcal F})$ to
  $\pi^*(\bar{\mathcal F})$. Then the triple ${\mathbf F}=({\mathcal F},\,\bar{\mathcal F},\,f)$ is an object of the equivariant derived category
	$D_G(X)$, defined in \cite{Be-Lu}.

If  ${\mathbf F}$ is an object as above,
let $\bar
{\mathcal I}^{\bullet}$ and $
{\mathcal I}^{\bullet}$ be complexes of injectives on 
 $\bar X$ 
 and on ${ X}$ that represent $\bar{\mathcal F}$ and ${\mathcal F}$, respectively.  By Proposition \ref{P:AUXforH}, 
 $$R^q\nu_*\bar{\mathcal F}=H^q(\nu_*(\bar{\mathcal I}^{\bullet}))=H^q(\Gamma(X,\,{\mathcal I}^{\bullet}))=H^q(X,\,{\mathcal F}).$$ 
Hence, $R^q\nu_*\bar{\mathcal F}$ is the constant sheaf on $BG$
defined by $H^q(X,\,{\mathcal F})$.

The equivariant cohomology $H_G(X,\,{\mathbf F})$ of ${\mathbf
F}\in { D}_G(X)$, is by definition the cohomology $H(\bar
X,\,\bar{\mathcal F})$; that is, the cohomology  $H(BG,\, R\nu_*\bar{\mathcal F})$.


\begin{Prop}\label{ThmSpectral}
The spectral sequence $E_2^{pq}=H^p(BG)\otimes H^q(X,\,{\mathcal
F})$
  abuts to $H^{p+q}_G(X,\,{\mathbf F})$.
\end{Prop}
 {\it Proof.} Since $R^q\nu_*\bar{\mathcal F}$ is the constant sheaf $H^q(X,\,{\mathcal F})$,
  the $E_2$ term of the Leray-Serre spectral sequence associated to the fibration
$X\to\bar X\to BG$ is
 $$E_2^{pq}=H^p(BG,\,R^q\nu_*(\bar{\mathcal F}))=H^p(BG)\otimes
H^q(X,\,{\mathcal F}),$$
 and the sequence abuts to $H^{p+q}(\bar X,\,\bar{\mathcal F})= H^{p+q}_G(X,\,{\mathbf F})$.
 \qed
 \begin{Cor} If  $H ^q(X,\, {\mathcal F})=0$ for $q$ odd, then  
$$H^{p+q}_G(X,\,{\mathbf F})\simeq H^p(BG)\otimes
H^q(X,\,{\mathcal F}).$$
\end{Cor}

{\it Proof.}
As the cohomology $H^p(BG)$ vanishes when $p$ is odd,   then the differential operators of  spectral sequence  $E_2^{pq}=H^p(BG)\otimes H^q(X,\,{\mathcal
F})$ vanish and  spectral sequence collapses. 

\qed 


 We assume that $G$ is
the {\em torus} $T=(U(1))^k$.   
The $T$-equivariant cohomology with complex coefficients of a
point $H_T({\rm pt};\,{\mathbb C})=H(BT;\,{\mathbb C})$ can be identified to the
algebra ${\mathbb C}[{\mathfrak t}^*_{\mathbb C}]$,  of polynomials on the complexification ${\mathfrak
t}_{\mathbb C}$ of the Lie algebra of $T$. 

We   denote by $\Xi$ the multiplicative subset of ${\mathbb
C}[{\mathfrak t}^*_{\mathbb C}]$ consisting of the non-zero
polynomials, and let $S$   denote the fixed point set of the
$T$-action.
 From the localization theorem
\cite[Sect. 6]{G-K-M} one deduces that the restriction map defines
an isomorphism
 $$H_T(X;\,{\mathbf F})_{\Xi}\to H_T(S;\,{\mathbf F})_{\Xi}$$
between the the corresponding localization modules.

As  the action of $T$ on $S$ is trivial, $ET\times_T S=BT\times S$. Thus, for the particular
case of complex coefficients
\begin{equation}\label{F-free}
H_T(S;\,{\mathbf F})\simeq H(BT)\otimes_{\mathbb C} H(S;\,{\mathcal F}).
 \end{equation}
 Hence,
 \begin{equation}\label{Sigma}
 H_T(X; \,{\mathbf F})_{\Xi}\simeq{\mathbb C}(t^*_{\mathbb
 C})\otimes_{\mathbb C} H(S; \,{\mathcal F}),
  \end{equation}
   ${\mathbb C}(t^*_{\mathbb C})$ being the field of rational
 functions on ${\mathfrak t}_{\mathbb C}$. In other words,
 $ H_T(X;\,{\mathbf F})_{\Xi}$ is the result of the extension of scalars in $H(S;\,{\mathcal F})$ from ${\mathbb C}$
 to ${\mathbb C}(t^*_{\mathbb C})$.

\begin{Prop}\label{P:EquivFiniteFixed}
Given ${\mathcal F}$ a brane which belongs to $\mathfrak{Coh}^T$, if the fixed point set for the $T$ action is $\{x_1,\dots,x_r\}$, then
$$H_T(X,\,{\mathbf F})\simeq\bigoplus_{i=1}^r\big({\mathbb C}({\mathfrak t}^*_{\mathbb C})  \otimes{\mathcal F}_{x_i} \big).$$
\end{Prop}

\begin{Cor}\label{CorFiniteFixed} Under the hypotheses of Proposition \ref{P:EquivFiniteFixed}, if ${\mathcal G}$ is other object of $\mathfrak{Coh}^T$ such that $\bigoplus_{i=1}^r\big({\mathbb C}({\mathfrak t}^*_{\mathbb C})  \otimes{\mathcal F}_{x_i} \big)$ and 
 $\bigoplus_{i=1}^r\big({\mathbb C}({\mathfrak t}^*_{\mathbb C})  \otimes{\mathcal G}_{x_i} \big)$ are not isomorphic as ${\mathbb C}({\mathfrak t}^*_{\mathbb C})$-vector spaces, then ${\mathcal F}$ and ${\mathcal G}$ are inequivalent $T$-equivariant branes.
\end{Cor}

\smallskip


 \subsection{Equivariant charges.}\label{SubsectEqch}  The charge of a brane  ${\mathcal F}$ which is
 a locally free ${\mathcal O}$-module is an element
  of the cohomology of $X$ defined from certain characteristic clases of $X$ and ${\mathcal
  F}$  \cite{Aspin, Harvey, M-M}. When ${\mathcal F}$ is a coherent sheaf, to define the
  charge it is necessary to pass to a locally free resolution of ${\mathcal
  F}$; the existence of such resolutions is a well-known fact  if $X$
  is a smooth variety  \cite{Fulton1}. This property permits to extent the definition  
  to objects of the Grothendieck group  of $X$, when $X$ is a variety. 
   In this section, we study the equivariant versions of that process. 
  
  The triangulated category $D(\mathfrak{Coh}^G)$, of $G$-equivariant branes on $X$, has associated the corresponding Grothendieck group
  $K(D(\mathfrak{Coh}^G))$. By $K^G(X)$ we denote the Grothendieck group of the abelian category $\mathfrak{Coh}^G$. The map
  \begin{equation}\label{Grothen-Groups}
  [{\mathcal F}]\in K( D(\mathfrak{Coh}^G))\mapsto \sum_i(-1)^i[H^i({\mathcal F})]\in K^G(X),
   \end{equation}
   where $[Z]$ the equivalence class of the object $Z$, is an isomorphism of abelian groups. 
   
   Let  $K'^G(X)$ be the Grothendieck group
of $G$-equivariant locally free ${\mathcal O}$-modules on $X$
\cite{Segal,Thomason}. In particular, $K'^G({\rm pt})$ is the ring $R(G)$ of virtual
 representations of $G$. The tensor product defines a ring structure
on $K'^G(X)$, and  the tensor product of locally free sheaves by coherent sheaves gives to $K^G(X)$ the structure of module on the ring  $K'^G(X)$.

  Given a $G$-equivariant brane, it seems  natural to assign it a $G$-equivariant charge. When the brane is a locally free sheaf, that assignation can be carried out   by means of the respective $G$-equivariant  characteristic classes.   The resulting charge will be an element of the equivariant cohomology
  $H_G(X)$.
  
   As in the non equivariant setting, an appropriate choice of the characteristic classes will permit to extent the definition to the objects of $K'^G(X)$.    To define an equivariant charge for branes in $\mathfrak{Coh}^G$,
   it is necessary to consider the cases for which the Grothendieck groups $K'^G(X)$ and $K^G(X)$ are isomorphic. In this situation, the equivariant charges are defined for arbitrary $G$-equivariant branes through  the isomorphism (\ref{Grothen-Groups}).

 One says that   $X$ has the $G$-equivariant resolution property if any $G$-equivariant
 coherent ${\mathcal O}$-module is the quotient of a locally free $G$-module.
 In this case, the natural homomorphism $K'^G(X)\to K^G(X)$ between the
 Grothendieck groups is an  isomorphism.

 Thomason \cite{Thomason1}
  proved the $G$-equivariant resolution property 
  for actions of linear algebraic groups which act on smooth varieties.
According with this result,
from now on in this subsection, we assume that:
  \begin{itemize}
  \item The K\"ahler manifold $X$ is an algebraic manifold; that is, $X$ admits a complex analytic embedding as a closed
  submanifold of ${\mathbb C}P^N$, for some $N$. \\
  \item  $G$ is a linear algebraic group. \\
  \item The action of $G$ on  $X$ is algebraic.
   \end{itemize}
 Under these assumptions the smooth algebraic variety $X$ has the $G$-resolution
  property and,  hence,
  the  homomorphism $K'^G(X)\to K^G(X)$ between the algebraic Grothendieck groups is an  isomorphism.

  By the GAGA principle, given a $G$-equivariant coherent ${\mathcal O}$-module ${\mathcal F}$, it is algebraic and  coherent with
  respect the algebraic structure.  So,   ${\mathcal F}$  is the quotient of an algebraic locally free $G$-equivariant sheaf  ${\mathcal E}_0$ on $X$.
    Consequently, it is possible to  construct a resolution
 \begin{equation} \label{resolution}
  0\to{\mathcal E}_m\to{ \mathcal E}_{m-1}\to\dots\to{\mathcal E }_0\to {\mathcal F}\to 0,
   \end{equation}
  consisting of $G$-equivariant  locally free sheaves and $m\leq {\rm
  dim}\,X$ \cite{Thomason1}.


Given an algebraic $G$-equivariant vector bundle $V$ on $X$ of
rank $r$, by the equivariant splitting principle \cite[page
315]{G-G-K}, $V$ has $r$ Chern roots: $x_1\dots,x_r$, such that
the equivariant Chern class $c^G_j(V)=\sigma_j(x_1,\dots,x_r)$,
with $\sigma_j$ the $j$th elementary symmetric function.

One defines the equivariant Chern character of $V$ by the
following element of the equivariant cohomology of $X$
$${\rm ch}^G(V)=\sum_1^re^{x_i}\in H_G(X,\,{\mathbb Q}).$$

The equivariant Chern character is ``additive" with respect the
exact sequences of equivariant vector bundles; that is, if $0\to
V'\to V\to V''\to 0$ is an exact sequence,
 then
${\rm ch}^G({V})={\rm ch}^G({ V}')+{\rm ch}^G({ V}'')$. 
 Hence, it
admits an extension to the algebraic Grothendieck group
$K'^G(X)\simeq K^G(X)$
 of $G$ equivariant coherent sheaves.

 In particular,
given a $G$-equivariant brane ${\mathcal
 F}\in\mathfrak{Coh}^G$, the complex ${\mathcal F}$
 is quasi-isomorphic to a complex ${\mathcal E}_{\bullet}$,
 where ${\mathcal E}_i$ is the sheaf of sections of a $G$-equivariant holomorphic vector bundle
 $V_i$. Then we define
$$ {\rm ch}^G({\mathcal F}):= \sum_i (-1)^i{\rm ch}^G({
V}_i).$$

The constant map  $p:X\to{\rm pt}$ gives rise to a pushforward
homomorphism $p_*:K(D(\mathfrak{Coh}^G))\to R(G)$ which maps the class $[{\mathcal F}]$
  to the virtual representation
$$\sum_i(-1)^iR^ip_*{\mathcal F}= \sum_i(-1)^iH^i(X,\,{\mathcal F}),$$
  where the representations
 on the cohomologies are the ones of Proposition
 \ref{Proprephcoho}.

The Chern character does not commutes with the pushforward by
proper maps \cite[page 280]{Fulton1}, \cite{Borel-Serre} and the
equivariant one neither \cite{E-G}.
If $V\to X$ is
 an equivariant vector bundle on $X$,
 then the non commutativity of the Chern character with the pushforward $p_*$ is
expressed by the statement of the equivariant
Hirzebruch-Riemann-Roch theorem:
$${\rm ch(p_*(V}))=p_*\big({\rm ch}^G(
V)\,{\rm td}^G(X)\big),$$
 where ${\rm td}^G(X)$ is the
$G$-equivariant Todd  of the bundle $TX$, that  can be defined by
means of  the equivariant splitting principle \cite[page
317]{G-G-K}.
  The equivariant Hirzebruch-Riemann-Roch
 allows us to express the pushforward ${\rm ch(p_*(V}))$  in terms of equivariant characteristic classes.

\smallskip

Let ${\mathcal F}$ be a general  brane of $D(\mathfrak{Coh}^G)$. The
previous remarks lead us to associate with ${\mathcal F}$ the
following
equivariant cohomology class
  \begin{equation}\label{EquiCharge0}
 Q^G({\mathcal F}):= {\rm ch}^G({\mathcal F})\,{\rm td}^G(X),
  \end{equation}
 that can be considered as an {\it equivariant charge} of the brane ${\mathcal F}$.
 
 Different equivariant charges can be defined by means of other  $G$-equivariant forms on $X$, for example $\sqrt{{\rm td}^G(X)}$, etc.


\smallskip

\noindent
{\bf Brane on an equivariant subvariety.}
We will consider the natural  brane defined by a subvariety of
the algebraic variety $X$. Let ${\mathcal I}$ be an ideal sheaf of
${\mathcal O}$. We put
$$Z={\rm supp}({\mathcal O}/{\mathcal I}).$$
$Z$ is an analytic subvariety of $X$ and ${\mathcal F}:={\mathcal
O}/{\mathcal I}$ is a coherent sheaf on $X$, that can be
considered as the structure sheaf of $Z$.

Locally, on an open $U$, the sheaf ${\mathcal I}$ will be generated
by the holomorphic functions $f_1,\dots,f_r$. We will assume that
each function is $G$-invariant. Hence, $Z$ is a $G$-invariant
subvariety of $X$ and ${\mathcal O}/{\mathcal I}$ is an
equivariant coherent sheaf.

Let us assume that $f_1,\dots,f_r$ is a regular sequence of
functions and let $e_1,\dots,e_r$ be the canonical basis of ${\mathbb
C}^r$. We put
$${\mathcal E}_k:={\mathcal O}\otimes_{\mathbb C}\big( \wedge^k{{\mathbb
C}^r} \big).$$
 Then ${\mathcal E}_k\simeq{\mathcal O}^{\oplus \tilde k},$ where
 $\tilde k:=\binom {r} {k}$. The known Koszul complex \cite[page 687]{G-H} is
 the following
 $G$-equivariant locally free resolution of ${\mathcal O}/{\mathcal
 I}$
 \begin{equation}\label{Koszul}
 0\to{\mathcal E}_r\overset{\partial}{\rightarrow}{\mathcal
 E}_{r-1}\to\dots\to{\mathcal E}_1\overset{\partial}{\rightarrow}{\mathcal E}_0={\mathcal O}\overset{\rm proj.}\to{\mathcal F}
 \to 0,
  \end{equation}
 where
$$\partial(h\otimes e_{j_1}\wedge\dots\wedge
e_{j_k})=\sum_i(-1)^ihf_i  e_{j_1}\wedge\dots \hat{e}_{j_i}
\dots\wedge e_{j_k}.$$
 Since the $f_i$ are $G$-invariant, the operator $\partial $ is
 equivariant.

 We can use the equivariant locally free resolution (\ref{Koszul}) to define ${\rm ch}^G({\mathcal O}/{\mathcal I})$
 $${\rm ch}^G({\mathcal O}/{\mathcal I})=\sum_{k=0}^r(-1)^k{\rm ch}^G({\mathcal E}_k)=\big(\sum_{k=0}^r(-1)^k\tilde k\big)\,{\rm ch}^G({\mathcal O}).$$
Since $\sum_{k=0}^r(-1)^k\tilde k=(-1+1)^r=0$, we have the following proposition
\begin{Prop}\label{PropchO/I}
The $G$-equivariant charge of the the brane ${\mathcal O}/{\mathcal I}$ is zero.
\end{Prop}


\smallskip

\noindent
{\bf Index of the Dirac operator.}
In some particular cases, the   charge has a natural
interpretation
 in terms of the index  of an elliptic operator.
 The exterior bundle $\Lambda^*T^*X$  of $X$ with the connection
 induced by the Levi-Civita connection  and the standard
 Clifford multiplication  is a Dirac bundle (see \cite[page 114]{L-M}). This  bundle
 has associated the corresponding Dirac operator ${\mathsf D}$. If $G$ acts as a group of isometries
  of $X$, then ${\mathsf D}$ is a $G$-operator \cite[page 211]{L-M}; i.e. ${\mathsf D}$ is $G$-equivariant.

 Let us assume that $G$ is compact and that
  ${ V}$ is a $G$-brane consisting of a locally free 
  sheaf. The compactness of $G$ allows us to average over the group
 for obtaining $G$-invariant metrics on $X$ and $G$-invariant connections on ${ V}$. On the other hand,
 the tensor product of $(\Lambda^*T^*X) \otimes{ V}$ is a Dirac bundle (see \cite[page 122]{L-M}) and the corresponding Dirac operator $D$ is also $G$-equivariant,  by the $G$-invariance of the metric and the connection.
  As $D$ is elliptic, ${\rm ker} \, D$ and ${\rm coker} \, D$ are representations
  of $G$ of finite dimension,  since $X$ is compact.
   For $g\in G$ the virtual character $\chi(D)(g)$ of $D$ at $g$ is defined by
  $$\chi(D)(g)={\rm trace}(g|_{{\rm ker}\,D})-{\rm trace}(g|_{{\rm coker}\,D}).$$

By the equivariant index theorem  \cite[Chapter 8]{B-G-V}, we have the proposition.
\begin{Prop}\label{P:DiracVirtual}
 In a neighborhood of
   $0\in{\mathfrak g}:={\rm Lie}(G)$
  \begin{equation}\label{chi(D)}
   \chi(D)\circ{\rm exp}=\int_XQ^G({ V}).
   \end{equation}
\end{Prop}

 The value of $\chi(D)({\rm exp}(\xi))$, for $\xi\in{\mathfrak g}$,  can be calculated by the localization formula in
 equivariant cohomology. The result is the Atiyah-Segal-Singer
 fixed point formula \cite{B-G-V,L-M}.
Next, we will evaluate  the integral (\ref{chi(D)}) when $X$ is a toric variety.

\smallskip
\noindent
{\it Toric varieties.}
Let $\Sigma$ be a fan in $N={\mathbb Z}^r$, such that each cone in $\Sigma$ is generated by a subset of  a basis of ${\mathbb Z}^r$.  We will denote by $X$
the smooth toric variety defined by $\Sigma$  \cite{C-L-S,Fulton,Oda}.
We put $M:={\rm Hom}_{\mathbb Z}(N,\,{\mathbb Z})$
 and $T$ for the torus
$$T=N\otimes{\mathbb C}^{\times}={\rm Hom}_{\mathbb Z}(M,\,{\mathbb
C}^{\times}).$$


We denote by $X^T$ the   fixed point set  for the $T$-action. For $x\in X^T$,
 let $\nu_{i,x}\in 2\pi({\mathbb Z})^n$,$\;i=1,\dots,n$, be the weights of the isotropy representation
 of $T$ on the tangent space $T_xX$. The fixed points of the $T$-action are in bijective correspondence
with the $n$-dimensional cones in $\Sigma$ \cite[\textsection
3.2]{C-L-S}. If the point $x$ is associated with the cone
$\sigma$, then
  \begin{equation}\label{omegaip}
  \omega_{i,x}=\frac{\nu_{i,x}}{2\pi}
  \end{equation}
  are the generators of $\sigma^{\vee}\cap M$, where $\sigma^{\vee}$ is the dual cone of $\sigma$.

Let $V$ be a $T$-equivariant holomorphic vector bundle on $X$ of range $m$. The
equivariant splitting principle together with the fact that  $T$ acts trivially on $X^T$, permit us to express the restriction    of $V$ to $X^T$ as
 a direct sum of $T$-equivariant line bundles
 $$V|_{X^T}=\bigoplus_{j=1}^m{ L}_j,$$
 where action of $T$ on $V|_{X^T}$ will be defined   by $m$ weights $\varphi_j$.
 Thus,    the $T$-equivariant
 Chern class of ${ L}_j$ is given by (see \cite{G-G-K})
  \begin{equation}\label{cT1}
  c^T_1({ L_j})=c_1({ L}_j)+\frac{1}{2\pi}\varphi_j.
   \end{equation}
 The $T$-equivariant Chern character of $V|_{X^T}$ is
 \begin{equation}\label{chT}
 {\rm ch}^T(V|_{X^T})=\sum_{j=1}^m{\rm exp}(c_1^T( L_j)).
  \end{equation}

\begin{Prop}\label{PropFixedPoint}
Let $X$ be a toric manifold and 
${ V}$  be a holomorphic $T$-equivariant vector bundle on $X$.
Denoting by $\{\varphi_{j,x}\}_{j=1,\dots,m}$ the weights of the
representation of $T$ on the fibre of ${ V}$ at a fixed
point $x$, then  (\ref{chi(D)}) is equal to
 \begin{equation}\label{2-n}
\sum_{x\in X^T}\Big(\sum_{j=1}^m{\rm
e}^{\frac{1}{2\pi}\varphi_{j,x}}
\Big)\prod_{i=1}^n\Big(1-{\rm e}^{-\omega_{i,x}}     \Big)^{-1},
 \end{equation}
 where $X^T$ is the set of fixed points of $X$ for the $T$-action and the $\omega_{i,x}$ are defined in (\ref{omegaip}).
\end{Prop}
{\it Proof.} 
The localization theorem  in equivariant cohomology
\cite[\textsection 10.9]{G-S} allows us to calculate the value
(\ref{chi(D)}) as a sum of contributions of the connected
components of $X^T$. As $X^T$ is discrete, the localization formula 
adopts the following form
\begin{equation}\label{chi(D)1}
 \chi(D)\circ{\rm exp}=(2\pi)^n\sum_{x\in X^T}\frac{Q^T(V)(x)}{\prod_i\nu_{i,x}},
 \end{equation}
where the $\nu_{i,x}$ are the weights of the isotropy representation of $T$ at the fixed point $x$.

From (\ref{cT1}) and (\ref{chT}), it follows
$${\rm ch}^T( V){\big|_x}=\sum_{j=1}^m{\rm e}^{\frac{1}{2\pi}\varphi_{j,x}}.$$
Similarly (see \cite[page 230]{L-M}),
$${\rm td}^T(TX){\big|_x}=\prod_{i=1}^n \omega_{i,x}\Big( 1-{\rm e}^{-\omega_{i,x}}   \Big)^{-1}.$$
The proposition follows from (\ref{EquiCharge0}) together with
(\ref{omegaip}). \qed

Note that
 the contribution of the manifold  $X$ to (\ref{2-n}) is encoded in the $n$-dimensional
cones of the fan $\Sigma$.



\section{Appendix}\label{S:Appendix}
 In this section, we will prove Propositions \ref{P:Propiso} and \ref{enumerate}.
 Let $({\mathcal H},\,\lambda)$ be an object of  $\mathfrak{Mod}^G$ and $g\in G$.


With the notations introduced in Subsection \ref{SubSectEquivSheav},  $b^*{\mathcal H}$ is the sheaf associated with the presheaf
 $$W\mapsto {\mathcal P}(W):={\mathcal O}_{G\times X}(W)\otimes_{b^{-1}{\mathcal O}(W)}b^{-1}{\mathcal H}(W),$$
 $W$ being an open set of $G\times X$. Given an open subset $U\subset X$ and $g\in G$, we put
  $U_g:=\{g\}\times U\subset G\times X.$ Identifying ${\mathcal O}_{G\times X}(U_g)$ with ${\mathcal O}(U)$,
   one has the following isomorphism of ${\mathcal O}(U)$-modules
   \begin{equation}\label{identif1}
   {\mathcal P}(U_g)= {\mathcal O}_{G\times X}(U_g)\otimes_{{\mathcal O}(U)}{\mathcal H}(U)\to{\mathcal
   H}(U),\;\;\;\tilde f\otimes\tau\mapsto f\tau,
  \end{equation}
  with $f(x)=\tilde f(g,\,x)$.

 Similarly, the sheaf $\mu^*{\mathcal H}$ is associated to the
 presheaf ${\mathcal M}$, with
$$W\mapsto {\mathcal M}(W):={\mathcal O}_{G\times X}(W)\otimes_{\mu^{-1}{\mathcal O}(W)}b\mu^{-1}{\mathcal H}(W).$$
One has the isomorphism of ${\mathcal O}(gU)$-modules
   \begin{equation}\label{identif2}
  {\mathcal M}(U_g)={\mathcal O}_{G\times X}(U_g)\otimes_{{\mathcal O}(gU)}{\mathcal H}(gU)\to{\mathcal
  H}(gU),\;\;\;\Hat h\otimes\tau\mapsto h\tau
   \end{equation}
    with $h(gx)=\Hat h(g,\,x)$.
    

 The restriction to $U_g$ of the morphism of sheaves
$\lambda$ is denoted  $\lambda|_{U_g}$
\begin{equation}\label{alpha(U}
\lambda|_{U_g}:b^{*}{\mathcal H}(U_g)
\longrightarrow \mu^{*}{\mathcal H}(U_g).
\end{equation}
By the above identifications (\ref{identif1}) and
(\ref{identif2}), $\lambda|_{U_g}$ determines an isomorphism of
${\mathcal O}(U)$-modules
 \begin{equation}\label{alpha(U}
 \lambda|_{U_g}: {\mathcal H}(U) \stackrel{\sim}{\longrightarrow} {\mathcal
 H}(gU),
 \end{equation}

where the ${\mathcal O}(U)$-module structure of ${\mathcal
 H}(gU)$ is defined through the isomorphism (\ref{O-LgO}).
  Thus,  we have proved   Proposition \ref{P:Propiso}.


As a consequence of  Proposition \ref{P:Propiso}, there is an isomorphism of ${\mathcal O}_x$-modules
 \begin{equation}\label{mathcalHx}
  \lambda_{g,x}:{\mathcal H}_x\to {\mathcal H}_{gx},
 \end{equation}
 where the  ${\mathcal O}_x$-structure of ${\mathcal H}_{gx}$ is induced by the isomorphism (\ref{O-LgO}).
 In this way, the isomorphisms (\ref{asterisco1}) are recovered.

\smallskip

Until now,in this appendix, we have only used the   existence of    the isomorphism
$\lambda$. Next, we will exploit the cocycle condition. Both sides
of equation (\ref{cocycle2}) are isomorphisms between two objects
on the category of ${\mathcal O}_{G\times G\times X}$-modules.
 The cocycle condition means  the commutativity of
the following triangle
 \begin{equation}\label{triangleZ}
  \xymatrix{
 {\mathcal Z}_1\ar[dr]_{(m\times 1_X)^*(\lambda)}\ar[rr]^{p^*(\lambda)} &&{\mathcal Z}_{2}\ar[dl]^{(1_G\times \mu)^*(\lambda)}\\
 & {\mathcal Z}_{3}, }
  \end{equation}
where
$$\begin{aligned}\notag
&{\mathcal Z}_1:=p^*b^*({\mathcal H})=(m\times 1_X)^*b^*({\mathcal
H}),\;\;{\mathcal Z}_2:=p^*\mu^*({\mathcal
H})=(1_G\times\mu)^*b^*({\mathcal H}) \\ &{\mathcal Z}_3:=(m\times
1_X)^*\mu^*({\mathcal H})=(1_G\times\mu)^*\mu^*({\mathcal H}).
\end{aligned}$$

As a consequence of the cocycle condition, we have the
following lemma.

\begin{Lem}\label{Lema}
Let $g,h$ be elements of $G$ and $U$ an open set of $X$, then
$$\lambda|_{(gU)_h}\circ\lambda|_{U_g}=\lambda|_{U_{hg}}.$$
In particular,
$\lambda|_{X_h}\circ\lambda|_{X_g}=\lambda|_{X_{hg}}$.
\end{Lem}

{\it Proof.} We consider   the commutative diagram
(\ref{triangleZ})
and  restrict this triangle to $\{h\}\times U_g\subset G\times
G\times X$. The restriction of $(m\times 1_X)^*(\lambda)$ is  the
morphism
$${\mathcal Z}_1(\{h\}\times U_g )={\mathcal H}(U)\longrightarrow {\mathcal Z}_3(\{h\}\times U_g )={\mathcal
H}((hg)U)$$
induced by $\lambda$. Thus,  by (\ref{alpha(U}), the mentioned restriction is
 $\lambda|_{U_{hg}}$.

 The restriction of $p^{*}(\lambda)$ to  $\{h\}\times U_g$
 $${\mathcal Z}_1(\{h\}\times U_g )={\mathcal H}(U)\longrightarrow {\mathcal Z}_2(\{h\}\times U_g )={\mathcal
H}(gU)$$
is (\ref{alpha(U}).

Finally, we consider the restriction of
$(1_G\times\mu)^*(\lambda)$. It is the morphism
 $${\mathcal Z}_2(\{h\}\times U_g )={\mathcal H}(gU)\longrightarrow {\mathcal Z}_3(\{h\}\times U_g )={\mathcal
H}(h(gU))$$
 induced by $\lambda$, and according to (\ref{alpha(U}) it is $\lambda|_{(gU)_h}$.
Then the lemma follows from   the commutativity of (\ref{triangleZ}).
 \qed

\smallskip

Proposition \ref{enumerate} is a direct consequence the Lemma \ref{Lema}.


\end{document}